\documentclass[journal]{IEEEtran}

\ifCLASSINFOpdf

\else

\fi

\usepackage{booktabs}
\usepackage{float}
\usepackage{color,soul}
\usepackage{graphicx}
\usepackage{algorithm}
\usepackage{algpseudocode}
\usepackage{siunitx}
\usepackage[numbers]{natbib}
\usepackage{amssymb}
\usepackage{amsmath}
\usepackage[hidelinks]{hyperref}
\usepackage[switch]{lineno}
\usepackage{chemformula}
\usepackage{multirow}

\begin{document}

\title{An Intelligent Energy Management Framework for Hybrid-Electric Propulsion Systems Using Deep Reinforcement Learning}

\author{\IEEEauthorblockN{Peng Wu,
		Julius Partridge},
	Enrico Anderlini, 
	Yuanchang Liu and
	Richard Bucknall

	\thanks{
		
		This work is partially supported by Royal Society (Grant no. IEC\textbackslash NSFC\textbackslash 191633). The authors thank Jens Christian Bjeldorf and Molslinje A/S for approving using the ship data in this study. The authors are grateful to Stig Eriksen and his colleagues for collecting the ship data. The authors are also indebted to Konrad Yearwood for his valuable critique of this article. \textit{(Corresponding author: Peng Wu.)}
		
		P. Wu, E. Anderlini, Y. Liu and R. Bucknall are with Department of Mechanical Engineering, University College London, London, WC1E 7JE, UK (emails: peng.wu.14@ucl.ac.uk, e.anderlini@ucl.ac.uk, yuanchang.liu@ucl.ac.uk, r.bucknall@ucl.ac.uk).
		
		J. Partridge is with AqualisBraemar LOC, Ibex House, 42-47 Minories, London, EC3N 1DY, UK (email: julius.partridge@abl-group.com).
}}

\maketitle

\begin{abstract}
	Hybrid-electric propulsion systems powered by clean energy derived from renewable sources offer a promising approach to decarbonise the world's transportation systems. Effective energy management systems are critical for such systems to achieve optimised operational performance. However, developing an intelligent energy management system for applications such as ships operating in a highly stochastic environment and requiring concurrent control over multiple power sources presents challenges. This article proposes an intelligent energy management framework for hybrid-electric propulsion systems using deep reinforcement learning. In the proposed framework, a Twin-Delayed Deep Deterministic Policy Gradient agent is trained using an extensive volume of historical load profiles to generate a generic energy management strategy. The strategy, i.e. the core of the energy management system, can concurrently control multiple power sources in continuous state and action spaces. The proposed framework is applied to a coastal ferry model with multiple fuel cell clusters and a battery, achieving near-optimal cost performance when applied to novel future voyages.
\end{abstract}


\IEEEpeerreviewmaketitle

\section{Introduction}
\label{sec:intro}

\subsection{Background and motivation}

Hybrid-electric propulsion systems comprising multiple power sources powered by clean energy derived from renewable sources are a promising approach to decarbonising the world's transportation systems \cite{smith2016co2, wu2020decarbonising, gray2021decarbonising}. An effective Energy Management System (EMS) is critical for hybrid-electric systems to manage power flows from, to and within the individual power sources, especially since such systems are typically constrained by high costs and will be subjected to operating under uncertainties \cite{sulaiman2018optimization}. These hybrid-electric systems are being extensively studied for rail \cite{peng2020scalable}, marine \cite{wu2020cost} and road \cite{wu2018stabilised} applications, focusing on developing effective EMS to achieve optimal or near-optimal performance. 

\subsection{Aim}

This article aims to propose a generic energy management framework for hybrid-electric propulsion systems. The energy management framework is based on Deep Reinforcement Learning (DRL) to deal with the high-dimensional control of multiple power sources operating under uncertainties, extending the authors' previous work in \cite{wu2020decarbonising, wu2020cost}. The proposed energy management framework is demonstrated through a case study of a typical coastal ferry with continuous monitoring of its operational status \cite{eriksen2018improving}. The DRL agent adopted in this study is the Twin Delayed Deep Deterministic Policy Gradient (TD3) \cite{fujimoto2018addressing} which is trained using an extensive volume of real-ship load profiles in multi-cluster fuel cell control settings. The energy management strategies generated by the TD3 agent are validated using novel load profiles collected over a period different from that during which the profiles used for training were recorded.

\subsection{Literature review}

EMS for hybrid-electric propulsion systems can be grouped into rule-based, optimisation-based and learning-based strategies \cite{tran2020thorough}. \citet{hofman2007rule} developed a rule-based EMS to minimise equivalent consumptions for hybrid vehicles. \citet{padmarajan2015blended} proposed a blended rule-based EMS for a plug-in hybrid-electric vehicle. \citet{peng2017rule} proposed an EMS calibrated by dynamic programming to achieve optimal performance for typical load profiles. Examples of rule-based EMS developed recently can be found in \cite{wang2019rule, ali2018optimized}. Although the development of rule-based EMS is relatively straightforward and can be easily implemented to guide the online controller of a hybrid-electric system, the method has limited resolution and it would be difficult to formulate the rules for very complex load profiles.

Optimisation-based EMS includes offline and online strategies. Offline optimisation-based EMS typically requires \textit{a priori} knowledge from typical load cycles. Consequently, they can only deliver the desired performance within the specific load profiles used to calibrate the EMS. Nevertheless, offline optimisation-based EMS can be used as a benchmark to evaluate the effectiveness of other online EMS, as long as there was \textit{a priori} knowledge of the complete load profile beforehand \cite{wu2020cost}. In contrast, an online strategy, such as Equivalent Consumption Minimisation Strategies (ECMS) and Model Predictive Control (MPC), does not require \textit{a priori} knowledge of the load profile and is typically formulated as an instantaneous optimisation problem for implementation with limited computational resources in real-time operations \cite{tran2020thorough}. \citet{kalikatzarakis2018ship} developed an ECMS-based online EMS for a hybrid-electric ship propulsion system. Their simulation results suggested an additional 6\% fuel saving can be achieved with ECMS for several load profiles. 
\citet{wang2016model} presented an MPC-based EMS for a hybrid-electric tracked vehicle, which achieved 6\% improvement in fuel economy over rule-based EMS for a few typical load profiles. 
It appears that optimisation-based online EMSs such as ECMS and MPC have achieved near-optimal performance when applied to a limited number of load profiles. However, their actual long-term performance remains unclear, especially for applications such as ships with highly scholastic load profiles \cite{wu2020decarbonising}. 

For a learning-based EMS, a model can be trained using an extensive volume of historical data to make real-time predictions of future power demands. Another model fits a complex control function derived by some other offline approaches (e.g. dynamic programming) that are computationally expensive for real-time applications \cite{murphey2012intelligentA, murphey2012intelligentB}. Although such an approach is valid, the fitted control function originally derived from dynamic programming is computationally expensive and is constrained by the `curse of dimensionality' \cite{sutton2018reinforcement}, making it impracticable for systems with high-dimensional and/or continuous state and actions spaces. In contrast, a Reinforcement Learning (RL) agent learns an optimal or near-optimal energy management strategy by continuously interacting with its `Environment' (including both the hybrid system model and historical load profiles) \cite{wu2020cost}. Effort from \cite{xiong2018reinforcement} solved the optimal energy management problem of a hybrid-electric vehicle using RL in discrete state and action spaces, with their RL agents trained with a limited number of load profiles. \citet{wu2020cost} trained their RL agent using large-scale continuous monitoring data, addressing the problem of function over-estimations when the environment is highly scholastic.  \citet{wu2018continuous} extended the state space to be continuous while the action space is still discrete. It is worth noting that all the above mentioned RL-based EMS were designed to control single power sources.

From the above analysis, it is evident that for hybrid-electric systems with simple load profiles, both conventional and novel learning-based EMS can deliver optimal or near-optimal energy management performance. However, applications such as ships typically have very complex and stochastic load profiles and require high redundancy within their propulsion systems with multiple power sources. There remains a challenge to derive an EMS that can offer concurrent control for systems similar to those found in ships that are required to manage multiple power sources operating under highly scholastic load profiles.

\subsection{Contributions}
In \cite{wu2020decarbonising}, the optimal energy management problem was solved in continuous state but discrete action spaces using DQN and Double DQN agents. Although both algorithms achieved voyage cost performance close to that of the off-line strategy solved by dynamic programming, such algorithms are limited to small discrete action space \citep{mnih2015human}. Considering fuel cell power level as a continuous parameter, this article extends the discrete action space to be continuous. In addition, in marine propulsion systems, especially Integrated Full Electric Propulsion (IFEP), for redundancy considerations, it is usual to install multiple power sources that can each be independently controlled. Therefore, this article will also explore the feasibility of controlling multiple fuel cell clusters using DRL algorithms.

\subsection{Organisation}
Section \ref{sec:problem} formulates the optimal energy management problem for hybrid-electric propulsion systems with multiple fuel cell clusters. Section \ref{sec:agent} introduces the Twin Delayed Deep Deterministic Policy Gradient (TD3) deep reinforcement learning algorithm \citep{fujimoto2018addressing}. Section \ref{sec:training} details the training process of the agent. Section \ref{sec:results} assesses the energy management strategy performance using unseen load profiles. Section \ref{sec:discussion} discusses the practical impacts of this work. Section \ref{sec:conclusions} concludes this article.

\section{Optimal energy management problem formulation}
\label{sec:problem}

\subsection{Candidate ship and the data}

The ship performance data used in this study is obtained from a coastal ferry operating between two fixed ports \cite{wu2020hybrid, eriksen2018improving}. It is intended that the plug-in hybrid fuel cell and battery propulsion system will replace the original diesel-based propulsion system, which has a power capacity of \SI{4370}{\kilo\watt} (five diesel generator sets, with each prime mover rated at \SI{874}{\kilo\watt}). The annual operating duty is 300 days, and the ship operates between two fixed ports accomplishing 16 voyages per day, with each crossing being of \SI{60}{\minute} duration \cite{wu2020hybrid}. It is assumed that battery charging can be carried out in both ports of the defined route, and hydrogen replenishment will take place overnight but never during the operational period \cite{wu2020cost}.

\subsection{Hybrid-electric propulsion system model}
\label{sec:hybrid_electric_system}

\begin{figure}
	\centering
	\includegraphics[width=1\linewidth]{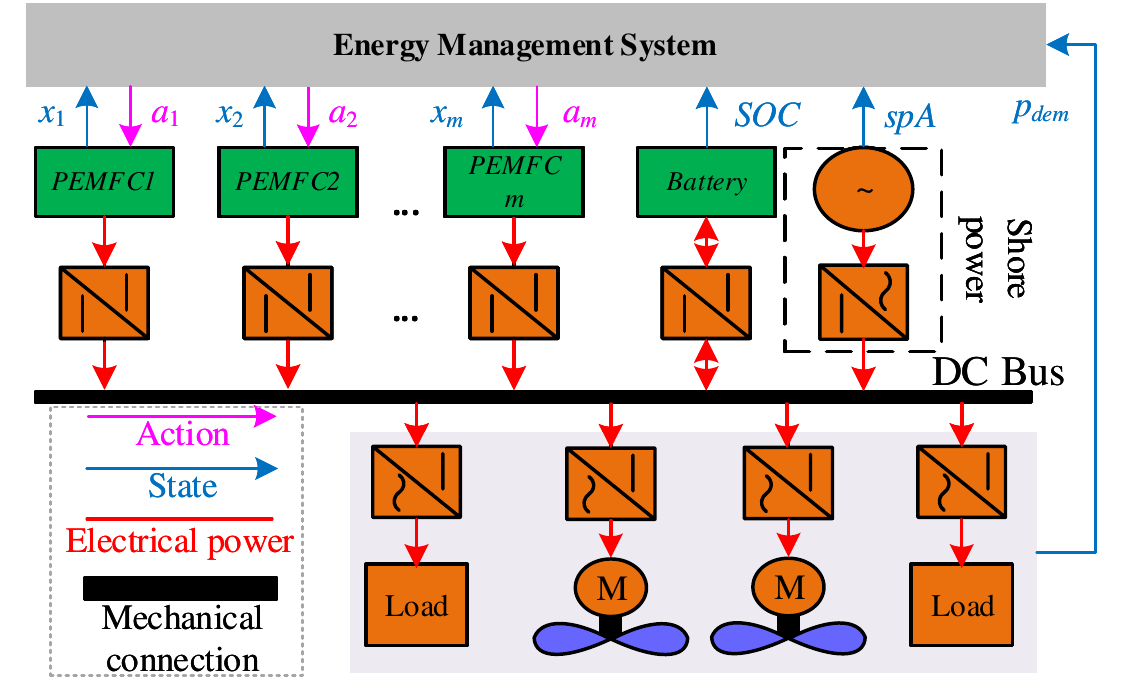}
	\caption{Hybrid-electric system model comprising multiple PEMFCs, battery and shore power connection.}
	\label{fig:hybrid_electric}
\end{figure}

Figure \ref{fig:hybrid_electric} provides an overview of the plug-in hybrid Proton Exchange Membrance Fuel Cell (PEMFC) and battery propulsion system model, which has been developed and optimised using the methodologies proposed in \cite{wu2018design} and \cite{wu2020hybrid}. This model has also been used in the authors' previous work for EMS development \cite{wu2020cost}. Readers may refer to \cite{wu2020cost} for more details. The model consists of a PEMFC operating on \ch{H2}, a lithium-ion battery, shore electricity supplies, an EMS, power converters and details of the total system power demand including both the propulsion and service loads. Note that the PEMFC rated maximum power output (2940 kW), maximum available battery capacity (581 kWh) and other model settings are identical as in \cite{wu2020cost}. In sailing mode, the battery can be charged by the fuel cell when excess power is available from the system. The battery is charged to its upper State of Charge (SOC) limit before sailing commences by shore-generated electricity while in port. For each time step, the model outputs the cost incurred in this time step:
\begin{equation}
	c_{t} = c_{b} + c_{f} + c_{h} + c_{e}
\end{equation}
where $c_{b}$, $c_{f}$, $c_{h}$ and $c_{e}$ are the costs incurred by battery degradation, fuel cell degradation, \ch{H_2} and shore-generated electricity consumption, respectively.

\subsection{RL-based EMS}

The objective of the EMS is to minimise costs incurred during operation of the hybrid-electric system. The optimal energy management problem of a hybrid-electric propulsion system can be modelled as a Markov Decision Process (MDP) \citep{wu2020cost}. For a system operating within a finite horizon $T$, the optimal $Q$ function is \citep{sutton2018reinforcement}:
\begin{equation}
	\label{eq:Q}
	Q^{*}(s,a) =
	\mathop{\mathbb{\max}}_{\pi} \mathbb{E}
	\left[\sum_{k=0}^{T}\gamma^{k}r_{t+k}|s_t = s, a_{t}=a\right]
\end{equation}
which leads to an optimal energy management strategy (i.e. the core of the EMS). Finding an optimal energy management strategy is to find an optimal policy $\pi^{*}$:
\begin{equation}
	\pi^{*}(s) = \operatorname*{arg\,max}_a \mathbb{E}\left[\sum_{k=0}^{T}\gamma^{k}r_{t+k}|s_t = s, a_{t}=a\right]
\end{equation}
where $t$ denotes current time step, $k$ is the number of time steps from $t$ to $T$, $\gamma \in  [0, 1]$ is the discount rate and the reward $r_{t}$ is a measure of the cost-effectiveness of taking action $a_{t}$ in state $s_{t}$ with the corresponding next system state $s_{t+1}$.

\begin{figure}
	\centering
	\includegraphics[width=1\linewidth]{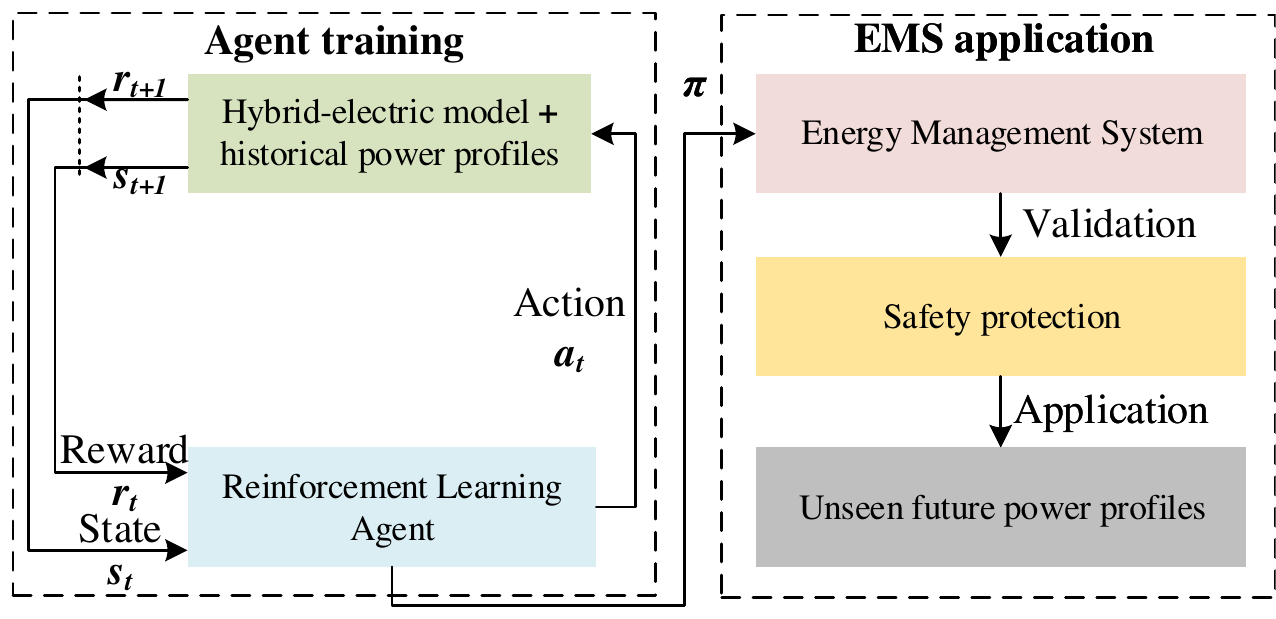}
	\caption{Schematic of the agent-environment interaction and EMS application.}
	\label{fig:agent_environment}
\end{figure}

Figure \ref{fig:agent_environment} illustrates the schematic the agent-environment interaction framework proposed to solve the MDP using RL and the procedure of applying the generated EMS. The environment comprises the plug-in hybrid PEMFC and battery propulsion system model and historical load profiles. The RL agent interacts with the environment by controlling the fuel cell power output and observing the reward signal and the resulting new system state.

\subsection{Action space}

The action is defined as the adjustment of fuel cell cluster per unit power ($[0, 1]$). In the preceding studies \cite{wu2020cost}, the fuel cell stacks are controlled uniformly in one-dimensional discrete action space. In this study, the action space is extended to be multi-dimensional and continuous to control multiple PEMFC clusters concurrently. Instead of controlling each PEMFC stack independently, PEMFC stacks are grouped into clusters, and stacks within one cluster are controlled uniformly. Such a setting simplifies the problem by avoiding very high dimensional action and state spaces. For the energy management problem with $m$ PEMFC clusters, the action space is defined as:
\begin{equation}
	a = 
	\left[\begin{matrix}
		a_1 	&
		a_2 	&
		\hdots	&
		a_{m-1} &
		a_m
	\end{matrix}\right]
\end{equation}
where $a_{k} \in \left[a_{M-},a_{M+}\right]$ is the $k^\mathrm{th}$ ($k = 1,2, \dots,m-1,m $) PEMFC cluster per unit power adjustment. Note that $a_{M-} = -0.04$, $a_{M+} = +0.04$ are maximum allowed per unit power decrease and increase limits, respectively. Note that all the fuel cell stacks are under uniform control when $m=1$.

\subsection{State space}
It is assumed that all PEMFC clusters are assigned with equal power output, and the rated cluster power is $P_{c} = \frac{P_\mathrm{fc,rated}}{m}$, where $P_\mathrm{fc,rated}$ is the total installed fuel cell power (see \cite{wu2020hybrid}). As each of the fuel cell clusters has its own power state, the fuel cell state is extended to:
\begin{equation}
	x = 
	\left[\begin{matrix}
		x_{1} 	&
		x_{2} 	&
		\hdots	&
		x_{m-1} &
		x_{m}
	\end{matrix}\right]
\end{equation}
where $x_{k}$ is the $k^\mathrm{th}$ fuel cell per unit power. The definitions of shore power availability $spA$, battery state of charge $SOC$, and power demand $p_\mathrm{dem}$ remain unchanged. Consequently, the new state space of the multi-stack energy management problem is:
\begin{equation}
	s = 
	\left[\begin{matrix}
		x_{1} 	&
		x_{2} 	&
		\hdots	&
		x_{m-1} &
		x_{m}  	&
		SOC  	&
		spA 	&
		p_{dem}
	\end{matrix}\right]
\end{equation}
With given action $a$, from current state $s$ to next state $s'$, the state transition is calculated by:
\begin{equation}
	s' =
	\!\begin{aligned}
		&
		\left[\begin{matrix}
			x_{1} + a_{1}		&
			x_{2} + a_{2} 		&
			\hdots		  		&
			x_{m-1} + a_{m-1} 	&\\
		\end{matrix}\right.\\
		&\qquad\qquad
		\left.\begin{matrix}
			x_{m} + a_{m}       & 
			SOC' 				&
			spA'				&
			p_{dem}'
		\end{matrix}\right]
	\end{aligned}
\end{equation}
where the next battery SOC, $SOC'$, is calculated using the system model \cite{wu2020hybrid} and the next shore power availability $spA'$ and $p_{dem}'$ is determined from the load profile. Note that the fuel cell power override function still applies \cite{wu2020cost}. It is worth mentioning that when calculating the battery output power, the total fuel cell power after the power converters, i.e. $P_{1}$ \cite{wu2020hybrid}, is updated to:
\begin{equation}
	P_{1} = \sum_{k=1}^{m} P_{c} x_{k} \eta_{1,k}
\end{equation}
where $\eta_{1,k}$ is the uni-directional power converter efficiency of the $k^\mathrm{th}$ PEMFC cluster.

\subsection{Reward function}

Based on the reward function described in \cite{wu2020decarbonising}, as multiple PEMFC clusters are configured, the reward function is updated to:
\begin{equation}
	r_{t+1} =
	\begin{cases}
		-1,    &   $spA=0$, \text{if } s_{t+1} \text{ is infeasible}\\
		-1, & $spA=0$,  \text{if any } x_{i}+a_{i,t} \notin [0,1] \\
		\displaystyle \tanh\left(\frac{1}{cost_{t+1}}\right),& $spA=0$, \text{else}\\
		\displaystyle \sum_{n=t+1}^{K} \tanh\left(\frac{1}{cost_{n}}\right), & $spA=1$\\
	\end{cases}
\end{equation}
where, when shore power is available ($spA=1$), the environment returns a summed reward of all the costs incurred in port mode of the current episode. In sailing mode, i.e. $spA=0$, the environment returns \num{-1} if $s_{t+1}$ is infeasible or one or more PEMFC cluster control actions are overridden. An episode is typically terminated at the last time step of a voyage. However, an episode would be terminated early if the battery's SOC is allowed to reduce to below a certain set limit or over-discharged. Note that the degradation and \ch{H2} fuel costs for each PEMFC cluster are calculated independently using the scalable PEMFC model, then subsequently summed to calculate the cost $cost_{t+1}$ incurred in one time step.

\section{Twin Delayed Deep Deterministic Policy Gradient}
\label{sec:agent}

Although Q-network based DQN and Double DQN agents delivered satisfactory performance in \cite{wu2020decarbonising}, these algorithms cannot be applied to problems with continuous or large action spaces due to the maximisation operation in selecting the optimal actions:
\begin{equation}
	a_t\gets
	\operatorname*{arg\,max}_a\left[Q\left(s_{t+1}, a;\theta \right)\right],
\end{equation} 
i.e. finding the greedy policy at every time step. To solve this problem, \citet{silver2014deterministic} proposed an actor-critic based Deterministic Policy Gradient (DPG) algorithm. With DPG, the actor $\pi_{\phi}$ (i.e. the policy), parametrised by $\phi$ is updated by taking the gradient of the expected return $J(\phi)$ \citep{silver2014deterministic, fujimoto2018addressing}:
\begin{equation}
	\label{eq:policygradient}
	\nabla_{\phi} J(\phi) = \mathbb{E}_{s \sim \rho^{\pi}} \left[\nabla_{a} Q^{\pi}\left(s,a\right)|_{a = \pi(s)} \nabla_{\phi}\pi_{\phi}(s) \right],
\end{equation}
where $Q^{\pi}\left(s,a\right) = \mathbb{E}_{s \sim \rho_{\pi}, a \sim \pi} \left[R_t |s,a \right]$ is the expected return of performing action $a$ in state $s$ following policy $\pi$. Note that $\rho^{\pi}$ is the state distribution which also depends on the policy parameters. $R_t = \sum_{i=t} ^T \gamma^{i-t} r(s_i,a_i)$, with a discount factor $\gamma$, is the discounted sum of rewards which the agent aims to maximise. Such an architecture is applicable to continuous control problems \citep{silver2014deterministic}. 

Later work of \citep{lillicrap2015continuous} developed a Deep Deterministic Policy Gradient (DDPG) based on DPG. In DDPG, exploration noise $\mathcal{N}_t$ is added to the actor policy $\pi_{\phi}(s_t)$:
\begin{equation}
	a_{t} = \pi_{\phi}(s_t) + \mathcal{N}_t
\end{equation}
The concept of experience replays and target networks have also been included in DDPG to break sample correlations and to improve training stability. The critic, i.e. the Q-function $Q_{\theta}(s,a)$ parametrised by $\theta$, is updated by minimising the loss:
\begin{equation}
	L = \displaystyle \frac{1}{D} \sum_{j=1}^D \left(y_{j} - Q_{\theta}(s_j,a_j)\right)^2
\end{equation} 
The policy $\pi_{\phi}$ is updated using the sampled policy gradient (see Eq. \ref{eq:policygradient}). In every training step, both the actor and critic target networks are soft-updated with a soft-update rate $\tau$.

\begin{algorithm}[H]
	\small
	\caption{Twin delayed policy deterministic policy gradients (TD3) agent \citep{fujimoto2018addressing}.}
	\label{alg:td3}
	\centering
	\begin{algorithmic}[1]
		\State{Initialise replay memory $D$ to capacity $M$}
		\State{Initialise critic networks  $Q_{\theta_{1}}$, $Q_{\theta_{2}}$, and actor network $\pi_{\phi}$with random parameters $\theta_{1}$, $\theta_{2}$, $\phi$ }
		\State{Initialise target networks $\theta_{1}' \gets \theta_{1}$, $\theta_{2}' \gets \theta_{2}$, $\phi' \gets \phi$}
		\While{$n<N_{max}$}
		\State{Initialise initial state $s_1$}    
		\For{$t = 1:T$}
		\State{Select action $a_{t}$ with exploration noise $a_{t}  \sim \pi_{\phi}(s_{t}) + \varepsilon$, $\varepsilon \sim \mathcal{N} \left(0, \sigma \right)$ }
		\State{Take action $a_{t}$, observe $r_{t+1}, s_{t+1}$ and $termination flag$}
		\State{Store transition $\left(s_{t}, a_{t}, r_{t+1}, s_{t+1} \right)$ in replay memory}
		\State{Every $Z$ steps sample random mini-batch of transitions $\left(s, a, r, s' \right)$ from mini-batch}
		\State{$\tilde{a} \gets \pi_{\phi'}\left(s' \right) + \tilde{\varepsilon}$, $\tilde{\varepsilon} \sim \text{clip} \left(\mathcal{N} \left(0, \tilde{\sigma}  \right),  -c, c \right)$}
		\State {Set $y =\begin{cases}
				r, & \text{if episode terminates}\\
				r + \gamma \min_{i=1,2} Q_{\theta_{i}}' \left(s', \tilde{a} \right), & \text{otherwise}\\
			\end{cases}$}
		\State{Update critics $\theta_{i} \gets \operatorname*{arg\,min}_{\theta_{i}} L_{\theta_{i}}$}
		\If{$t \text{ mod } d= 0$}
		\State{Update $\phi$ by the deterministic policy gradient:}
		\State{$\nabla_{\phi} J\left(\phi \right) = D^{-1} \sum \nabla_a Q_{\theta_{1}} \left(s, a\right) |_{a = \pi_{\phi}(s)} \nabla_{\phi} \pi_{\phi}(s) $}
		\State{Soft-update target network:}
		\State{$\theta'_{i}  \gets \tau \theta_{i} + \left(1 - \tau \right) \theta_{i}' $}
		\State{$\phi'  \gets \tau \phi_{i} + \left(1 - \tau \right) \phi' $}
		\EndIf
		\State{Terminate if  $termination flag$}
		\EndFor
		\EndWhile
	\end{algorithmic}
\end{algorithm}

Although DDPG can deliver satisfactory performance in some continuous control tasks, the overestimation bias can be problematic \citep{fujimoto2018addressing}. \citet{fujimoto2018addressing} proposed the Twin Delayed Deep Deterministic Policy Gradient (TD3) (Algorithm~\ref{alg:td3}) addressing the function approximation errors and overestimation bias in actor-critic methods. The results in \cite{wu2020cost} and \cite{wu2020decarbonising} have highlighted the problem of overestimations in stochastic environments. Consequently, a novel approach using the TD3 algorithm is proposed in this work to solve the optimal energy management problem of the plug-in hybrid PEMFC/battery system.

In TD3, there are two critic networks, i.e. $Q_{\theta_{1}}$ and $Q_{\theta_{2}}$ and one actor network $\pi_{\phi}$. The subscripts of $\theta_{1}$, $\theta_{2}$ and $\phi$ denote the neural network parameters. Correspondingly, there are two critic target networks $Q'_{\theta_{1}}$  and $Q'_{\theta_{2}}$ and one actor target network $\pi'_{\phi}$. The superscript $'$ denotes target network. Three key improvements have been made to DDPG \citep{fujimoto2018addressing}.

The first improvement is clipped Double Q-learning for the actor-critic. When calculating the target value, the minimum value between the two critic estimates is selected:
\begin{equation}
	y \gets r + \gamma \min_{i = 1,2} Q'_{\theta_{i}}\left(s', \tilde{a} \right)
\end{equation}
such that the less biased Q-value estimate is used, which is similar to the concept of Double Q-learning \citep{NIPS2010_3964}. Note that $\tilde{a}$ is given by the target actor network with a small amount of random noise added to the target action to smooth the value estimate by bootstrapping off of similar state-action value estimates (second improvement) \citep{fujimoto2018addressing}:
\begin{equation}
	\tilde{a} \gets \pi_{\phi'}\left(s' \right) + \tilde{\varepsilon}, \tilde{\varepsilon} \sim \text{clip} \left(\mathcal{N} \left(0, \tilde{\sigma}  \right),  -c, c \right)
\end{equation}
where $\tilde{\varepsilon} \in \left(-c, c\right)$ is the added noise clipped from a Gaussian distribution $\mathcal{N} \left(0, \tilde{\sigma}  \right)$  with a \num{0} mean value and a standard deviation of $\tilde{\sigma}$ such that the target is maintained close to the original action. Moreover, the actor $\pi_{\phi}$ parametrised by $\phi$ is updated less frequently than the critics, i.e. $\phi$ is updated every $d$ critic updates such that accumulated errors can be reduced.

It is worth noting that the Huber loss function is adopted to update the critic networks, which is different from the Mean Squared Error used in \citep{fujimoto2018addressing}. This adjustment is made to achieve more stable agent training as the Mean Squared Error loss function in \cite{wu2020decarbonising} leads to diverged training processes. Consequently, the loss function for the critic is:
\begin{equation}
	L_{i} \left( \theta \right) = \displaystyle \frac{1}{D} \displaystyle \sum_{j=1} ^{D} \sigma_{i,j}
\end{equation}
where:
\begin{equation}
	\sigma_{i,j} = 
	\begin{cases}
		\displaystyle \frac{1}{2} \delta_{i,j} ^2, & \text{if } \left| \delta_{i,j} \right|  < 1 \\
		\displaystyle \left| \delta_{i,j} \right| - \displaystyle \frac{1}{2}, & \text{otherwise}
	\end{cases}
\end{equation}
where $\delta_{i,j} = y_{i,j} - Q_{\theta_{i}}(s,a)$, $i=1,2$, $j = 1,2,\dots, D$. $\delta$ denotes temporal difference. $i$ denotes the $i^\mathrm{th}$ critic; $j$ denotes the $j^\mathrm{th}$ sample in the mini-batch with capacity $D$.

\section{Agent training}
\label{sec:training}
For multi-cluster fuel cell control, i.e. the fuel cells are distributed to multiple clusters and are controlled separately. The cluster number $m$ is set to 4. This results in the number of total power sources being five (four PEMFC clusters and one battery). This is to provide a hybrid configuration that mirrors the original IFEP configuration with 5 diesel generators. It is worth noting that, when $m=1$, the fuel cells are controlled uniformly as in \cite{wu2020cost} but in continuous action space. Only the multi-cluster ($m=4$) results are presented in this article. Readers may refer to \cite{wu2020decarbonising} for other relevant results and the hyperparameter settings.

The agents were trained on a workstation with two Intel Xeon E5-2683 V3 processors (28 cores in total). The environment and the agent were coded in Python. The agent's neural networks were built and trained with PyTorch v1.20. 

As the learning curve of the agent can be influenced by the random seeds (determining the appearance order of training load profiles to the agent) of the environment \citep{henderson2018deep} each agent was trained with 28 different random seeds for reproducibility. The agent policy performance was assessed by calculating the average values and standard deviations across all instances where convergence was achieved. Note that as the neural networks are relatively small, only one CPU thread is assigned to each running instance to avoid training speed degradation due to unnecessary parallelisation. 

The actual strategy performance was tested periodically (every 100 training episodes) using 10 random training voyages. Note that in test mode, no exploration noise was added; and battery over-discharge protection was enabled (disabled in training mode). Once the training of all the 28 instances was completed, the agent with the lowest episode cost was selected to generate detailed energy management strategy results. 

\subsection{Neural network settings}

\begin{figure}
	\centering
	\includegraphics[width=1\linewidth]{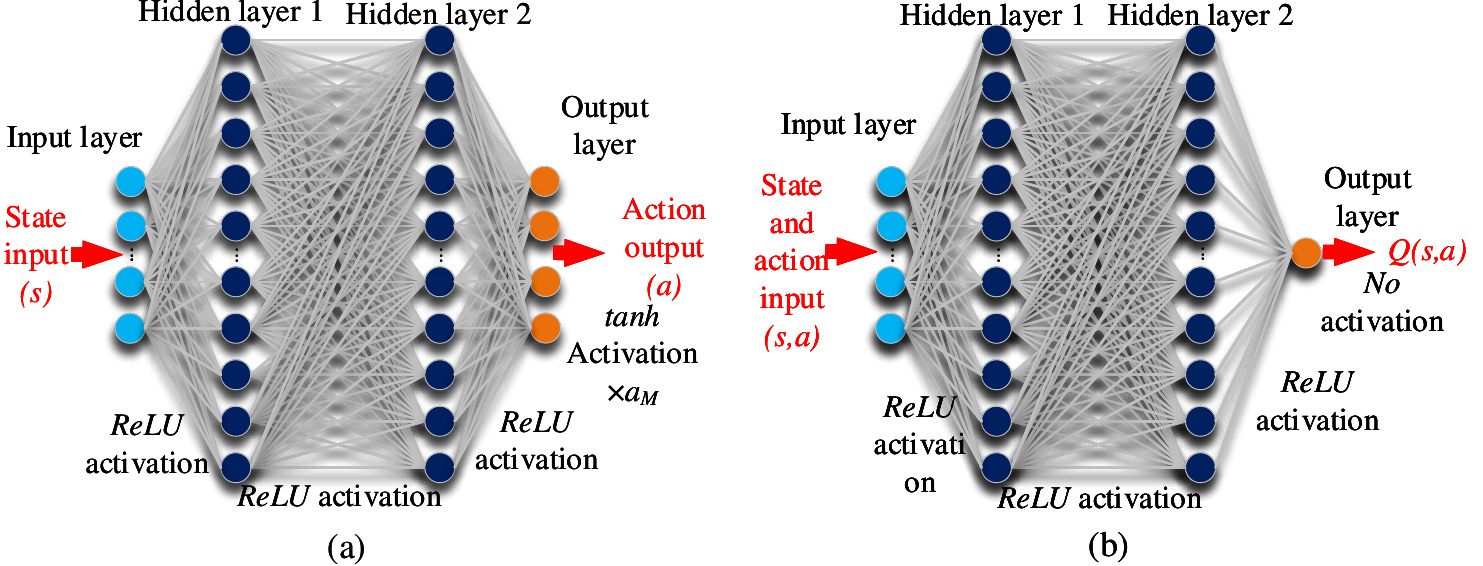}
	\caption[Neural network settings of TD3.]{Neural network settings of TD3. (a) Actor network setting and (b) Critic network setting. Each hidden layer has 256 neurons.}
	\label{fig:network}
\end{figure}

Figure \ref{fig:network} illustrates the settings for the actor (a) and critic (b). The actor observes state inputs $s$ and chooses action $a$.  As in Figure \ref{fig:network}a, the inputs to the actor are state vectors. Two fully-connected hidden layers forward propagate the state inputs followed by a fully-connected output layer. The input layer and the two hidden layers are activated by ReLU. The output layer is activated by a hyperbolic tangent function (tanh) then multiplied by $a_M$ to match the action limits of $[-0.04, 0.04]$. As in Figure \ref{fig:network}b, the critic receives the state and action inputs, and outputs the Q-value (see Section \ref{sec:agent}). Note that there is no activation function applied to the output layer to allow free value estimates. For the 4-cluster control, the state and action space dimensions are 7 and 4, respectively.

\subsection{Training}

\begin{figure}
	\centering
	\includegraphics[width=1\linewidth]{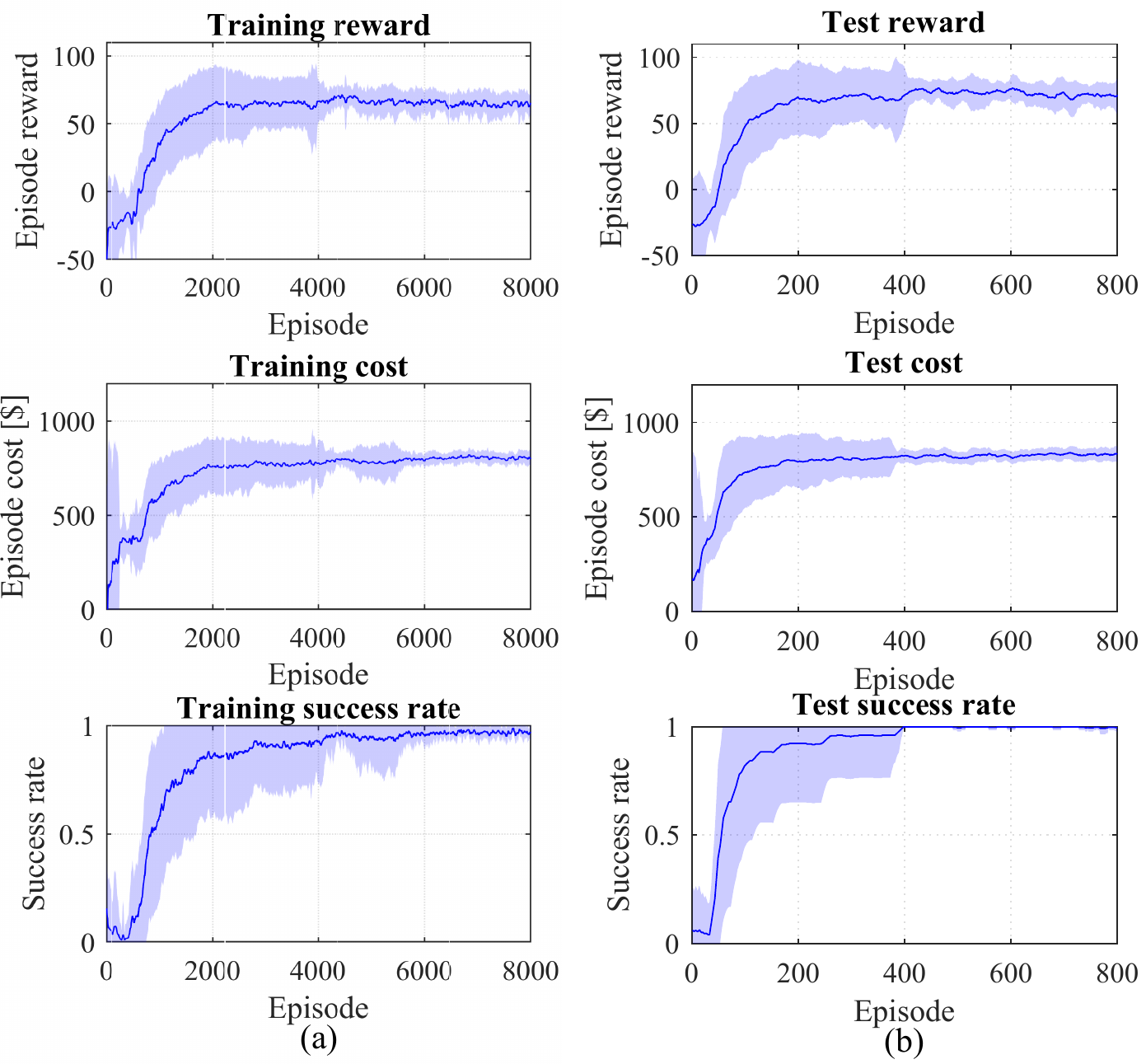}
	\caption[4-cluster fuel cell control training and testing with Huber loss function.]{4-cluster fuel cell control training and testing with Huber loss function. The deep blue lines are moving average values across 26 converged instances running with different random seeds. Two diverged instances are not included. The light blue shadows are the confidence bounds calculated by mean values $\pm$ standard deviations across the 26 instances.}
	\label{fig:multi_trainingandtest}
\end{figure}

Figure \ref{fig:multi_trainingandtest} illustrates the training process for the 4-cluster fuel cell control EMS. The training was terminated at \num{8000} episodes. 2 out of the 28 instances diverged. It required \SI{204}{\minute} for a converged instance to complete the training. The voyage cost converged to a value slightly higher than \$800.0.

\section{Results}
\label{sec:results}

\subsection{Overview}

The learned strategy was first applied to the training voyages to verify the EMS cost performance without any policy noise. As the learned strategy is intended to achieve minimum voyage cost for un-predicted future voyages, the EMS was validated by applying to a set of un-predicted future voyages. 

As each of the strategies was generated with different random seeds vary slightly from the others, the strategy with minimum average voyage costs was selected to generate the detailed voyage power distributions. The training and validation voyage sets are detailed in \cite{wu2020cost} and \cite{wu2020decarbonising}. The load profile samples are also identical to the ones discussed previously.

In this section, the energy management strategy is applied to validation voyages which were not included in the training dataset such that the EMS performance for future unknown voyages can be assessed.

\subsection{Validation sample 1 with low power demand}

\begin{figure}
	\centering
	\includegraphics[width=1\linewidth]{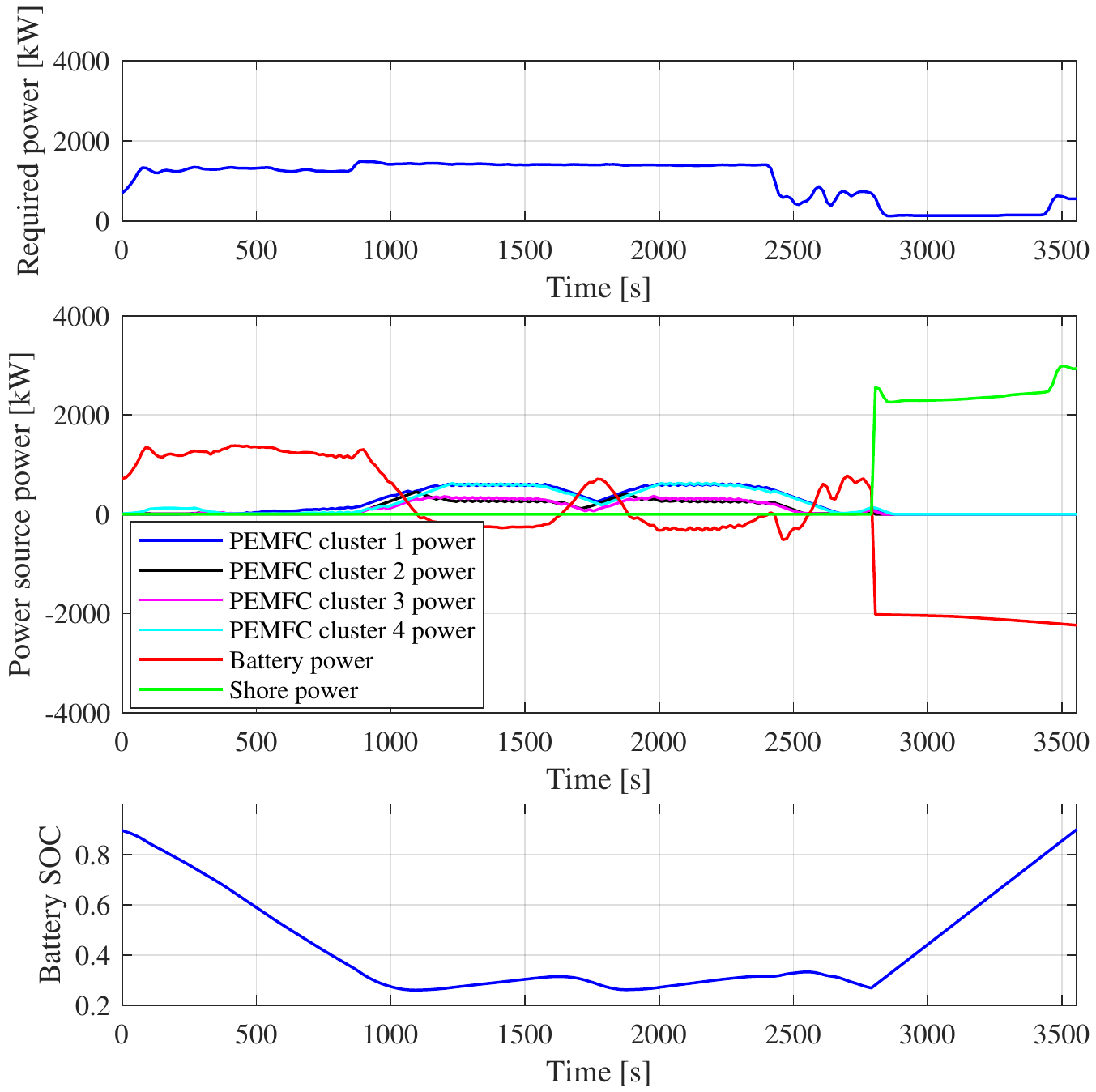}
	\caption{TD3 4-cluster strategy for validation sample voyage 1 with low power demand.}
	\label{fig:validation_light_multi}
\end{figure}

Figure \ref{fig:validation_light_multi} presents the TD3 4-cluster strategy for validation sample voyage 1. Clusters 1 and 4 show similar trajectories with higher loads. The power output of clusters 2 and 3 are also similar but are lower compared to clusters 1 and 4. Due to early fuel cell starts and unnecessary power adjustments, the TD3 4-cluster strategy leads to a 20.1\% higher PEMFC degradation cost as depicted in Table  \ref{tab:multi_single_validation1}. The voyage Global Warming Potential (GWP) emission of the TD3 4-cluster strategy is lower by 2.1\%  due to reduced electricity and \ch{H2} consumption.

\begin{table}
	\centering
	\footnotesize
	\caption{Comparison of TD3 uniform and multi-cluster strategy voyage costs and GWP emissions for validation sample 1.}
	\label{tab:multi_single_validation1}
	\begin{tabular}{@{}ccccccc@{}}
		\toprule
		\multirow{3}{*}{} & \multicolumn{3}{c}{Voyage cost} & \multicolumn{3}{c}{Voyage GWP Emission} \\
		& Multi & Uniform & $\frac{\text{Multi}}{\text{Uniform}}$ & Multi & Uniform & $\frac{\text{Multi}}{\text{Uniform}}$ \\
		& [\$] & [\$] & [\%] & [kg] & [kg] & [\%] \\ \midrule
		PEMFC & 259.4 & 216.1 & 120.1 & - & - & - \\
		Battery & 63.7 & 63.7 & 100.0 & - & - & - \\
		Electricity & 43.7 & 44.0 & 99.2 & 81.7 & 82.3 & 99.2 \\
		\ch{H2} & 347.4 & 361.2 & 96.2 & 63.2 & 65.8 & 96.2 \\
		\textit{Sum} & 714.3 & 685.1 & 104.3 & 144.9 & 148.1 & 97.9 \\ \bottomrule
	\end{tabular}%
\end{table}

\subsection{Validation sample 2 with moderate power demand}

\begin{figure}
	\centering
	\includegraphics[width=1\linewidth]{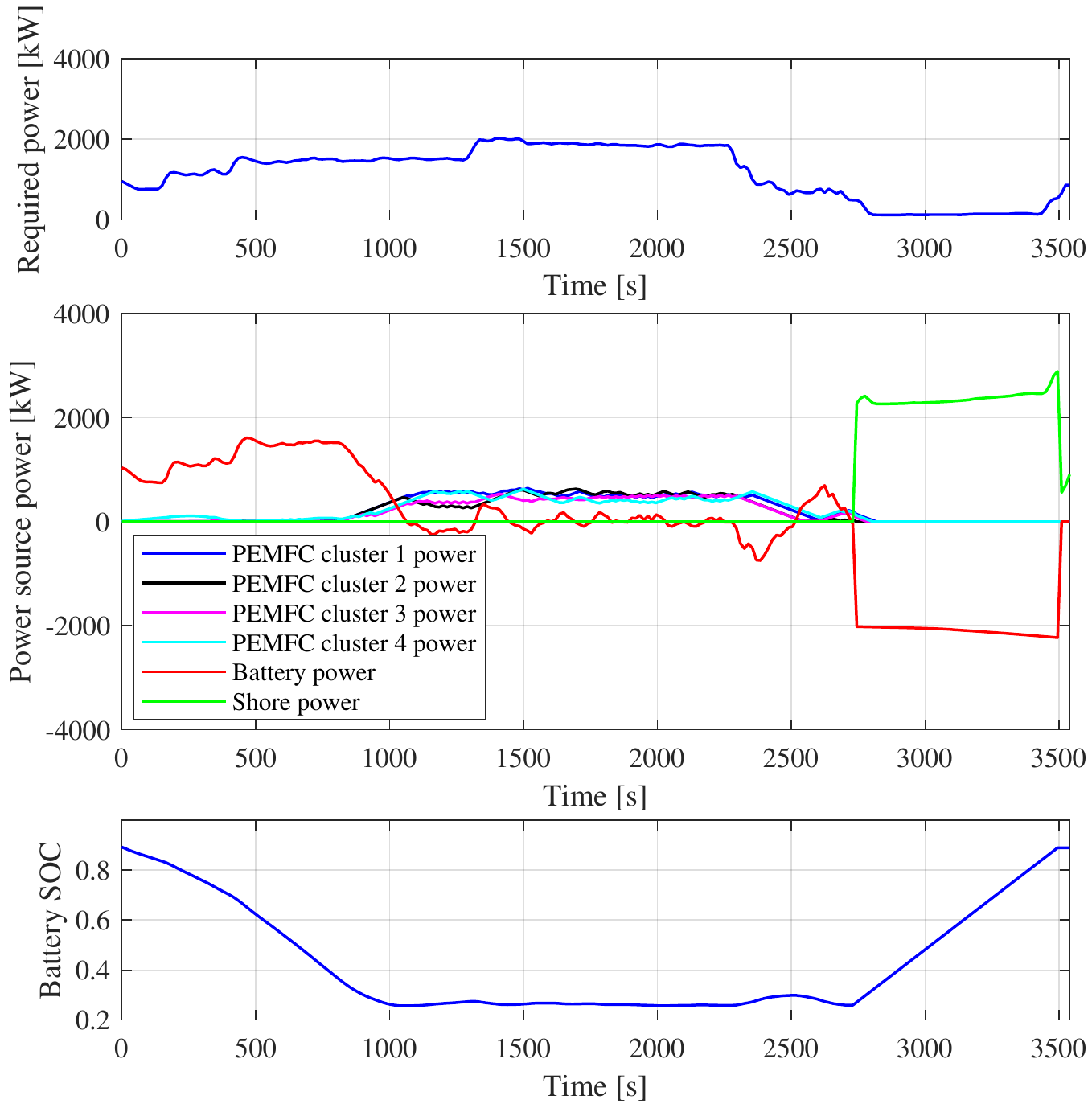}
	\caption{TD3 4-cluster strategy for validation sample voyage 2 with moderate power demand.}
	\label{fig:validation_moderate_multi}
\end{figure}

Figure \ref{fig:validation_moderate_multi} shows the TD3 4-cluster strategy for validation sample 2. As the PEMFC cluster outputs are adjusted frequently, the PEMFC degradation cost of the 4-cluster strategy is 19.6\% higher than that of the uniform strategy (Table \ref{tab:multi_single_validation2}). Although the 4-cluster strategy increases both electricity and \ch{H2} costs, the overall voyage cost of the 4-cluster strategy is 3.8\% lower than that of the uniform strategy. Owing to increased electricity and \ch{H2} consumption, the 4-cluster strategy produces 1.4\% higher voyage GWP emissions.

\begin{table}
	\centering
	\footnotesize
	\caption{Comparison of TD3 uniform and multi-cluster strategy voyage costs and GWP emissions for validation sample 2.}
	\label{tab:multi_single_validation2}
	\begin{tabular}{@{}ccccccc@{}}
		\toprule
		\multirow{3}{*}{} & \multicolumn{3}{c}{Voyage cost} & \multicolumn{3}{c}{Voyage GWP Emission} \\
		& Multi & Uniform & $\frac{\text{Multi}}{\text{Uniform}}$ & Multi & Uniform & $\frac{\text{Multi}}{\text{Uniform}}$ \\
		& [\$] & [\$] & [\%] & [kg] & [kg] & [\%] \\ \midrule
		PEMFC & 204.7 & 244.8 & 83.6 & - & - & - \\
		Battery & 63.7 & 63.7 & 100.0 & - & - & - \\
		Electricity & 44.9 & 44.7 & 100.5 & 84.0 & 83.6 & 100.5 \\
		\ch{H2} & 437.9 & 427.6 & 102.4 & 79.7 & 77.8 & 102.4 \\
		\textit{Sum} & 751.3 & 780.8 & 96.2 & 163.7 & 161.4 & 101.4 \\ \bottomrule
	\end{tabular}%
\end{table}

\subsection{Validation sample 3 with high power demand}

\begin{figure}
	\centering
	\includegraphics[width=1\linewidth]{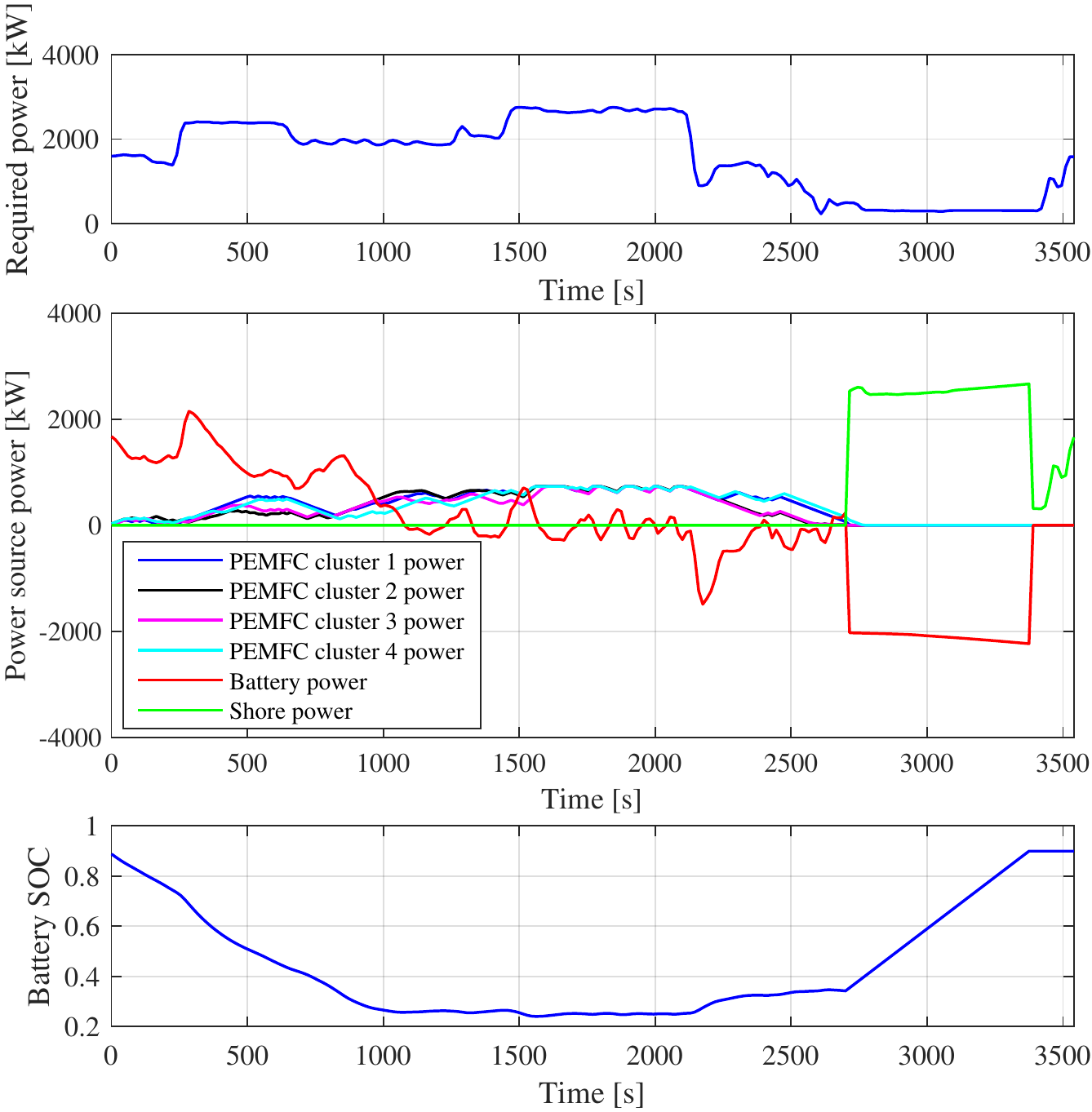}
	\caption{TD3 4-cluster energy management strategy for validation sample voyage 3 with high power demand.}
	\label{fig:validation_heavy_multi}
\end{figure}

Figure \ref{fig:validation_heavy_multi} illustrates the TD3 4-cluster strategy for validation sample voyage 3. Although the battery handles most of the large power transients, the PEMFC clusters are occasionally adjusted, leading to a higher PEMFC degradation cost (Table \ref{tab:multi_single_validation3}). The voyage cost of the 4-cluster strategy is 5.7\% higher than that of the uniform TD3 strategy due to the increase in \ch{H2} consumption and PEMFC degradation. The voyage GWP emission of the 4-cluster TD3 strategy is 3.0\% lower as a result of reduced \ch{H2} consumption.

\begin{table}
	\centering
	\footnotesize
	\caption{Comparison of TD3 uniform and multi-cluster strategy voyage costs and GWP emissions for validation sample 3.}
	\label{tab:multi_single_validation3}
	\begin{tabular}{@{}ccccccc@{}}
		\toprule
		\multirow{3}{*}{} & \multicolumn{3}{c}{Voyage cost} & \multicolumn{3}{c}{Voyage GWP Emission} \\
		& Multi & Uniform & $\frac{\text{Multi}}{\text{Uniform}}$ & Multi & Uniform & $\frac{\text{Multi}}{\text{Uniform}}$ \\
		& [\$] & [\$] & [\%] & [kg] & [kg] & [\%] \\ \midrule
		PEMFC & 246.1 & 267.8 & 91.9 & - & - & - \\
		Battery & 63.7 & 63.7 & 100.0 & - & - & - \\
		Electricity & 49.1 & 48.9 & 100.4 & 91.8 & 91.4 & 100.4 \\
		\ch{H2} & 678.1 & 715.9 & 94.7 & 123.4 & 130.3 & 94.7 \\
		\textit{Sum} & 1037.1 & 1096.4 & 94.6 & 215.2 & 221.8 & 97.0 \\ \bottomrule
	\end{tabular}%
\end{table}

Table \ref{tab:compare_all} details the TD3 4-cluster strategy performance in comparison with the TD3 uniform, Double DQN, Double Q-Learning and the off-line optimal Deterministic Dynamic Programming (DDP) strategies. The 4-cluster strategy average voyage cost of the validation voyages is \$783.1, which is 2.7\% higher than that of the TD3 uniform strategy. The TD3 uniform strategy emits 1.8\% less GWP emissions for validation voyages. Nevertheless, both the TD3 uniform and 4-cluster strategies have delivered satisfactory performance when undertaking future novel voyages, in comparison with other on-line strategies. Note that the aim of multi-cluster control is improved system redundancy.

\begin{table}[]
	\centering
	\footnotesize
	\caption{Comparison of TD3 uniform and multi-cluster strategy voyage average costs and GWP emissions with other algorithms.}
	\label{tab:compare_all}
	\begin{tabular}{@{}ccccc@{}}
		\toprule
		\multirow{3}{*}{Algorithm} & \multicolumn{2}{c}{Average cost} & \multicolumn{2}{c}{Average GWP} \\
		& Cost & Ratio to DDP & GWP & Ratio to DDP \\
		& [kg] & [\%] & [\$] & [\%] \\ \midrule
		DDP & 723.5 & 100.0 & 158.4 & 100.0 \\
		Double Q-Learning & 813.8 & 112.5 & 149.2 & 94.2 \\
		Double DQN & 768.9 & 106.3 & 157.5 & 99.4 \\
		TD3 uniform & 762.5 & 105.4 & 163.3 & 103.1 \\
		TD3 4-cluster & 783.1 & 108.2 & 166.3 & 105.0 \\ \bottomrule
	\end{tabular}
\end{table}

\section{Discussion}
\label{sec:discussion}

It was observed that the multi-cluster TD3 strategy led to less than a 3\% average voyage cost increase compared to the uniform TD3 strategy. The reasons for higher voyage costs and GWP emissions are: (1) the multi-cluster strategy adjusts cluster power frequently which would lead to increased PEMFC degradations and (2) the multi-cluster strategy will, for extended periods, operate with one or more of the clusters at very low power settings leading to low fuel efficiency. Though finer hyperparameter tuning and more extended training might further improve the EMS cost and emission performance. Another possible cause could be the limitations of the algorithm when subjected to a high dimensional action space. These possibilities would require further investigation. Nevertheless, it should be noted that the multi-cluster control framework is intended to improve system redundancy. Consequently, the tiny deviations are acceptable when the actual strategy performance is near-optimal in comparison with the off-line strategy.

The energy management framework has been demonstrated through a coastal ferry case study to provide control references for its fuel cell clusters within its hybrid-electric propulsion system in both state and action spaces. For other applications, either road vehicles or ships, with similar hybrid-electric propulsion configurations, the proposed energy management framework can be adapted to achieve near-optimal operational objectives such as minimum costs or equivalence of emissions. Nevertheless, the overestimation issues need to be properly addressed to generate feasible strategies for unseen load profiles. This work also demonstrates how to use continuous monitoring data with deep reinforcement learning effectively.

\section{Conclusions}
\label{sec:conclusions}

This article aimed to propose a generic energy management framework for hybrid-electric propulsion systems. In this framework, the action space of the optimal energy management problem has been extended to be continuous and multi-dimensional to explore the feasibility of controlling multiple fuel cell clusters in a continuous action space using deep reinforcement learning. A generic multi-cluster fuel cell environment has been implemented based on the one developed in previous studies. A Twin Delayed Deep Deterministic Policy Gradient with Huber loss function has been applied to solve the updated energy management problem in multi-dimensional and continuous action space. As a special case of multi-cluster control, the uniform fuel cell control was first solved using TD3. The uniform strategy learned by the TD3 agent further reduces the average voyage cost in both training and validation load profiles. The novel multi-cluster fuel cell control framework that was developed can be used to achieve optimal control of multiple power sources in a stochastic environment. By updating the system model, load profiles and the deep reinforcement learning agents, the developed energy management framework can be adapted to other hybrid-electric propulsion systems to achieve objectives such as minimising operational costs and equivalent emissions for future deployments.

\ifCLASSOPTIONcaptionsoff
\newpage
\fi

\bibliographystyle{IEEEtranN}
\bibliography{referencesIEEE}

\end{document}